\documentclass[aps,pra, preprint,amsmath,amssymb,amsfonts,superscriptaddress]{revtex4-2}

\usepackage{graphicx}       
\usepackage{dcolumn}        
\usepackage{bm}             
\usepackage[version=3]{mhchem} 
\usepackage{xcolor}         
\usepackage[normalem]{ulem} 
\usepackage{lipsum}         
\usepackage[colorlinks=true, allcolors=blue]{hyperref}
\usepackage{amsmath}
\usepackage{physics}
\usepackage[font=small]{caption}

\usepackage{subcaption}

\begin{document}

\title{Quantum dynamics of an ion in an RF quadrupole trap with weak octupolar anharmonicity}

\author{Sukanya Jana}
\affiliation{School of Chemical Sciences, Indian Association for the Cultivation of Science, Kolkata-700032, India}

\author{Anushree Dey}
\affiliation{School of Chemical Sciences, Indian Association for the Cultivation of Science, Kolkata-700032, India}

\author{Pintu Mandal*}
\affiliation{Department of Physics, St. Paul’s Cathedral Mission College, Kolkata-700009, India}

\author{Nabanita Deb*}
\affiliation{School of Chemical Sciences, Indian Association for the Cultivation of Science, Kolkata-700032, India}
\email{nabanita.deb@iacs.res.in}

\date{\today}

\begin{abstract}

We investigate the quantum dynamics of a single ion confined in a radio-frequency (RF) Paul trap in the presence of weak octupole-induced anharmonicity, as can occur in realistic ion-trap potential. The explicitly time-dependent quadrupole potential is treated in a reference-oscillator Fock basis, allowing the effects of the RF drive and the quartic correction to be resolved directly at the level of motional-state couplings. While the quadratic RF interaction generates $\Delta n=\pm2$ transitions and Gaussian squeezing, the quartic contribution opens additional $\Delta n=\pm4$ coupling pathways. Starting from the reference-oscillator vacuum, we show that the octupole interaction broadens the even-parity Fock-state distribution and enhances the phonon-number fluctuations more strongly than the mean occupation. At the same time, the RF-induced sub-vacuum quadrature squeezing remains robust, although its magnitude and orientation are modified by the quartic interaction. The number statistics and quadrature fluctuations therefore respond differently to the same nonlinear perturbation, reflecting the distinct roles of Fock-space redistribution and phase coherence. The resulting dynamics provide a quantum-mechanical characterization of weak octupole corrections beyond the ideal quadrupole approximation and clarify how realistic anharmonicities modify nonclassical trapped-ion motion. Importantly, the motional nonclassicality remains robust against the weak octupole perturbation, as evidenced by persistent sub-vacuum quadrature squeezing and Klyshko-inequality violation. These results characterize how realistic anharmonic corrections reshape trapped-ion motion beyond the ideal quadrupole approximation.

\end{abstract}

\maketitle
\section{Introduction}

Trapped atomic ions confined in radio-frequency (RF) Paul traps combine long coherence times, high-fidelity state preparation and readout, and precise control of both internal and motional degrees of freedom, making them a leading platform for quantum science and technology \cite{leibfried2003quantum,blatt2008entangled,haeffner2008quantum,bruzewicz2019trapped,brown2016co,wineland2011quantum,xu20253d}. Seminal proposals and experiments established trapped-ion quantum logic and scalable architectures \cite{cirac1995quantum,monroe1995logic,cirac2000scalable,home2009complete}, while the same controllability has enabled programmable quantum simulation of interacting spin and spin--boson systems \cite{blatt2012quantum,monroe2021programmable,foss2025progress,pagano2025fundamentals,schmied2011quantum}. Trapped ions also play a central role in precision spectroscopy, quantum-logic spectroscopy, and optical frequency standards \cite{ludlow2015optical,schmidt2005spectroscopy,rosenband2008frequency,diddams2001optical}. These applications all rely, directly or indirectly, on accurate control of the ion's quantized motion and therefore motivate a detailed understanding of motional dynamics within the trap.

The motional degree of freedom is not merely a source of systematic error; it is also a controllable quantum resource. Resolved-sideband cooling permits preparation near the motional ground state \cite{monroe1995cooling,king1998cooling}, and laser--ion interactions have been used to generate and manipulate Fock, coherent, squeezed, and Schr\"odinger-cat-like states of motion \cite{meekhof1996generation,cirac1994nonclassical,poyatos1996strong,blatt1995quantum}. State-reconstruction protocols and direct measurements of motional density matrices and phase-space distributions have made the oscillator state experimentally accessible \cite{leibfried1996experimental,poyatos1996tomography}. More recently, motional squeezing has been exploited as a resource for enhanced spin--spin interactions and for coherent transport and separation protocols \cite{ge2019squeezed,sutherland2021motional}. These developments connect trapped-ion motion naturally to the broader theory of Gaussian and non-Gaussian bosonic states \cite{weedbrook2012gaussian,adesso2014continuous}, while number-statistical criteria provide complementary diagnostics of nonclassicality \cite{klyshko1996observable}. Beyond the usual Lamb--Dicke and rotating-wave regimes, wave-packet studies have further shown that richer interference and revival structures can emerge in the motional phase space \cite{wang2008interference}.

At the level of the trap itself, the RF confinement makes the motional Hamiltonian intrinsically time periodic. The classical coordinate obeys a Mathieu equation, whereas its quantum counterpart is a harmonic oscillator with a periodically modulated spring constant. Exact and quasienergy-based treatments have established the quantum structure of this problem \cite{brown1991quantum,gheorghe1992quantum}, while invariant and propagator methods provide complementary descriptions of the corresponding time-dependent oscillator \cite{lewis1969exact,harari2011propagator}. Direct wave-packet calculations have clarified the relation between the full RF dynamics, micromotion, and effective time-independent descriptions \cite{hashemloo2016wave,hai2003wavepacket}, and quantum studies have also explored regular and nontrivial dynamical structures in Paul traps \cite{moore1993quantum}. More generally, periodically driven quantum systems are naturally organized in a Floquet framework \cite{shirley1965solution}, an approach that has recently become an active control tool in trapped-ion experiments, including engineered vibrational dynamics and periodically driven qubits \cite{kiefer2019floquet,he2021riemann}. Retaining the explicit RF dependence is therefore important whenever micromotion and drive-induced quantum correlations are themselves part of the physics of interest.

Real trapping potentials, however, are never perfectly quadrupolar. Finite electrode dimensions, machining or fabrication tolerances, and imperfect electrode geometries generate higher-order multipole components, while uncompensated fields and surface noise can introduce additional motional shifts and decoherence \cite{berkeland1998micromotion,brownnutt2015noise,labaziewicz2008suppression}. The influence of nonideal multipole fields in Paul traps has been recognized for decades \cite{dawson1969non}. In particular, octupole contributions produce quartic corrections to the trapping potential and lead classically to Duffing-like amplitude-dependent secular-frequency shifts \cite{sevugarajan1999field,sevugarajan2000frequency}. Anharmonic trapping potentials can also modify normal-mode frequencies and mode structure in ion crystals \cite{home2011normal}. Analyses of nonlinear Paul traps have shown that weak octupole fields alter stability boundaries and motional frequencies \cite{zhou2010characteristics}, and more detailed treatments have identified nonlinear harmonics and resonance phenomena generated by the added octupole component \cite{xiong2015theoretical,xiong2016nonlinear}. These effects become increasingly relevant when high motional precision is required or when miniaturized trap geometries enhance the relative importance of nonquadratic terms.

The quantum consequences of anharmonic confinement have consequently attracted attention in several settings. Anharmonic traps have been considered for the evolution of nonclassical motional states \cite{sharma2002cat}, for resolving motional transitions in proposed quantum-information encodings \cite{wang2012shor}, and for simulating elementary quantum dynamics through controlled motional-state transformations \cite{santos2015simulation}. Squeezed coherent-state descriptions have also been developed for quadrupole--octupole RF traps \cite{mihalcea2018squeezed}, while engineered anharmonic confinement has been proposed for scalable ion trap architectures \cite{lin2009large}. Related hybrid-system studies have shown that trap anharmonicities can qualitatively affect the driven quantum motion of an ion coupled to a nanomechanical degree of freedom \cite{fountas2019nanowire}. 

\textcolor{black}{Despite these developments, the interplay between the intrinsic RF periodicity of the quadrupolar confinement and a weak octupolar-induced quartic nonlinearity has not been explored systematically at the level of motional Fock-state dynamics. In particular, how the additional Fock-space couplings introduced by the quartic term redistribute population among motional states and modify their number and phase-sensitive quantum properties under explicitly time-dependent RF driving remains insufficiently characterized.}

\textcolor{black}{In the present work, we develop a time-dependent Fock-basis treatment of a single ion confined in an RF Paul trap with a weak octupolar contribution. Rather than replacing the RF field by a static pseudopotential, we retain the explicit periodic driving and incorporate the octupolar component through its RF-modulated quartic contribution. This formulation makes the distinct state-to-state couplings generated by the quadratic and quartic terms explicit: the former produces \(\Delta n=\pm2\) transitions, while the latter opens additional \(\Delta n=\pm4\) channels and introduces occupation-dependent diagonal shifts. We then characterize the resulting motional dynamics through Fock-state populations, phonon-number statistics, and phase-space measures of nonclassicality, including quadrature squeezing, Wigner distributions, and the Klyshko criterion. By comparing the ideal quadrupolar and weakly anharmonic cases, we establish how a physically motivated octupolar correction modifies the nonclassical motional state of the ion.}

The remainder of the paper is organized as follows. Section \ref{theoretical_framework} presents the theoretical formulation and Fock-space representation of the driven quadrupole--octupole trap. Section \ref{numerical_methods} describes the numerical methods and observables, while Sec. \ref{results_and_discussions} presents the results and discussion. The main conclusions are summarized in Sec.\ref{Conclusions}, with additional derivations provided in the Appendices \ref{mat_ele} -- \ref{app:marginal}.

\section{Theoretical Framework}\label{theoretical_framework}
\subsection{Formulation of Quantum Hamiltonian} \label{Formulation_QH}

The quadrupole potential in the radial plane of a linear Paul trap is
\begin{equation}
  \Phi_2(x,y,t) = \left(U + V_0\cos\Omega t\right)\frac{x^2-y^2}{r_0^2},
  \label{eq:Phi2}
\end{equation}
where $U$ and $V_0$ denote the DC and RF voltage amplitudes, respectively,
$\Omega$ is the angular frequency of the RF drive, and $r_0$ is the
characteristic trap radius. The classical motion of an ion of charge $e$ and
mass $m$ is governed by the Mathieu equation
\begin{equation}
  \frac{d^2u}{d\xi^2} + \left[a_u + 2q_u\cos(2\xi)\right]u = 0,
  \qquad u = x, y,
  \label{eq:Mathieu}
\end{equation}
where $\xi = \Omega t/2$ and the dimensionless Mathieu parameters are
\begin{equation}
  a \equiv a_x = -a_y = \frac{8eU}{m\Omega^2 r_0^2},
  \qquad
  q \equiv q_x = -q_y = \frac{4eV_0}{m\Omega^2 r_0^2}.
  \label{eq:aq}
\end{equation}
Restricting the analysis to the $x$ direction ($y=0$), the potential energy
$e\Phi_2$ of the ion in the ideal quadrupole field can be expressed in terms of
the Mathieu parameters as
\begin{equation}
  V_2(x,t) = \frac{m\Omega^2}{8}\left(a + 2q\cos\Omega t\right)x^2 .
  \label{eq:V2}
\end{equation}
The first term describes the static harmonic confinement, whereas the second
represents its periodic modulation by the RF drive.
 
In a realistic trap, deviations of the electrode geometry from the ideal
hyperbolic shape give rise to higher-order multipole contributions, among which the octupole term is of particular relevance. The
corresponding octupolar potential reads
\begin{equation}
  \Phi_4(x,y,t) = A_4\left(U + V_0\cos\Omega t\right)
  \frac{x^4 - 6x^2y^2 + y^4}{r_0^4},
  \label{eq:Phi4}
\end{equation}
where the dimensionless coefficient $A_4$ quantifies the strength of the
octupole relative to the quadrupole contribution and is fixed by the electrode
geometry. The octupole term introduces anharmonicity into the otherwise harmonic confinement and, through the $x^2y^2$ term, couples the two radial directions, which are decoupled in the ideal quadrupole approximation. Restricting the dynamics to the $x$ direction, the corresponding potential energy is
\begin{equation}
  V_4(x,t) = \frac{m\Omega^2 A_4}{8 r_0^2}\left(a + 2q\cos\Omega t\right)x^4 .
  \label{eq:V4}
\end{equation}
 
Because $\Phi_2$ and $\Phi_4$ are generated by the same electrode voltages, the
quadratic and quartic terms share the same time dependence. Combining
Eqs.~\eqref{eq:V2} and \eqref{eq:V4}, the total potential energy of the ion
along $x$ is
\begin{equation}
  V(x,t) = V_2(x,t) + V_4(x,t)
         = \frac{m\Omega^2}{8}\, f(t)\left(x^2 + \kappa\, x^4\right),
  \label{eq:Vtot}
\end{equation}
where we have introduced the dimensionless drive function $f(t) = a + 2q\cos\Omega t$ and the octupolar anharmonicity parameter $\kappa = A_4/r_0^2$.

The quantum dynamics follows from promoting the position and momentum to
Hermitian operators $\hat{x}$ and $\hat{p}$ satisfying
$[\hat{x},\hat{p}] = i\hbar$. The Hamiltonian of the ion then reads
\begin{equation}
  \hat{H}(t) = \frac{\hat{p}^2}{2m}
             + \frac{m\Omega^2}{8}\, f(t)\,\hat{x}^2
             + \kappa\,\frac{m\Omega^2}{8}\, f(t)\,\hat{x}^4 ,
  \label{eq:H}
\end{equation}
the anharmonicity is itself modulated by the RF drive,
whereas the ideal Paul trap corresponds to $\kappa = 0$.
 
It is convenient to express the dynamics in units set by the RF drive. We
introduce the oscillator length associated with the drive frequency, $\ell_\Omega = \sqrt{\hbar/m\Omega}$, together with the dimensionless position, momentum, and time, $\hat{X} = \frac{\hat{x}}{\ell_\Omega}$, $\hat{P} = \frac{\ell_\Omega}{\hbar}\,\hat{p}$ and $\tau = \Omega t$, which satisfy $[\hat{X},\hat{P}] = i$. Measuring energy in units of
$\hbar\Omega$, the Hamiltonian (Eq.~\ref{eq:H}) takes the form
\begin{equation}
  \hat{\mathcal{H}}(\tau) \equiv \frac{\hat{H}}{\hbar\Omega}
  = \frac{\hat{P}^2}{2}
  + \frac{f(\tau)}{8}\,\hat{X}^2
  + \lambda\,\frac{f(\tau)}{8}\,\hat{X}^4 ,
  \label{eq:Hdimless}
\end{equation}
with $f(\tau) = a + 2q\cos\tau$. \textcolor{black}{Here, $\lambda = \kappa\,\ell_\Omega^2$ = $A_4\,\ell_\Omega^2/r_0^2$ is a dimensionless parameter that characterizes the strength of the octupolar anharmonicity relative to the quadrupole contribution. Since the characteristic oscillator length $\ell_\Omega$ is comparable to the characteristic trap radius $r_0$, the perturbation parameter is of the order of the octupole coefficient, $\lambda\sim A_4$. The coefficient $A_4$ is determined by the electrode geometry and therefore characterizes the strength of the octupolar contribution arising from the specific trap configuration.}
 
Equation~\eqref{eq:Hdimless} describes a periodically driven anharmonic
oscillator in which, for the octupolar contribution considered here, both the harmonic and the quartic terms are modulated
by the RF drive. The dynamics is thus fully characterized by the three
dimensionless parameters $a$, $q$, and $\lambda$: the Mathieu parameters $a$
and $q$ determine the classical stability and the micromotion of the ion, while
$\lambda$ quantifies the octupolar anharmonicity arising from the electrode
geometry. For $\lambda \neq 0$, the quartic
term breaks this exact solvability and renders the motion intrinsically
nonlinear, so that the motional spectrum and dynamics acquire an
amplitude-dependent character absent in the ideal trap. The Hamiltonian (Eq. \ref{eq:Hdimless}) constitutes the starting point for the analysis presented
in the following sections.

\subsection{Fock Space Representation}

We expand the quantum state of the ion in the number-state (Fock) basis $\{\ket{n},\,n=0,1,2,\ldots\}$ of a reference harmonic oscillator with frequency $\omega_0=\sqrt{a}\,\Omega/2$. This choice corresponds to the static quadratic part of the trapping Hamiltonian, since $(1/2)m\omega_0^2\hat{x}^2=(m\Omega^2a/8)\hat{x}^2$. Thus, $\omega_0$ defines the reference Fock basis used throughout the calculation.

The time-dependent state is written as $\ket{\psi(t)}=\sum_{n=0}^{N-1}c_n(t)\ket{n}$, where $c_n(t)\in\mathbb{C}$ are the Fock-state amplitudes and $N$ is the Fock-space truncation. The normalization condition $\sum_{n=0}^{N-1}|c_n(t)|^2=1$ is preserved by the exact unitary dynamics and monitored numerically during the propagation.

Introducing the harmonic-oscillator scaling factors $x_{\rm sc}=\sqrt{\hbar/2m\omega_0}$ and $p_{\rm sc}=\sqrt{m\hbar\omega_0/2}$, the position and momentum operators can be written using the ladder operators $\hat a$ and $\hat a^\dagger$ as $\hat{x}=x_{\rm sc}(\hat{a}+\hat{a}^{\dagger})$ and $\hat{p}=ip_{\rm sc}(\hat{a}^{\dagger}-\hat{a})$.
With the choice $\omega_0=\sqrt{a}\,\Omega/2$, the kinetic energy and the static quadratic confinement combine exactly as 
\begin{equation*}
    \frac{\hat{p}^2}{2m}+\frac{m\Omega^2a}{8}\hat{x}^2=\hbar\omega_0\left(\hat{a}^{\dagger}\hat{a}+\frac{1}{2}\right).
\end{equation*} 
Consequently, the static quadratic contribution is diagonal in the chosen Fock basis, whereas the RF modulation proportional to $q\cos\Omega t\,\hat{x}^2$ generates off-diagonal couplings between the Fock states.


The selection rules follow directly from the non-zero matrix elements derived in Appendix \ref{mat_ele}. For the quadratic operators, the only off-diagonal elements that survive are of the form $\bra{n+2}\hat{x}^2\ket{n}$ and $\bra{n+2}\hat{p}^2\ket{n}$, together with their Hermitian conjugates. Thus, apart from the diagonal contributions, $\hat{x}^2$ and $\hat{p}^2$ directly couple Fock states differing by two oscillator quanta, giving $\Delta n=0,\pm2$.

For the quartic operator, in addition to the diagonal and $\Delta n=\pm2$ matrix elements, the elements $\bra{n+4}\hat{x}^4\ket{n}$ and their Hermitian conjugates are also non-zero. Consequently, the quartic interaction allows $\Delta n=0,\pm2,\pm4$ transitions. The $\Delta n=\pm4$ terms therefore provide additional direct coupling pathways that are absent in the purely quadratic trap.

Since all non-zero couplings connect Fock states whose quantum numbers differ by an even integer, the Hamiltonian preserves Fock-state parity. Equivalently, because the Hamiltonian contains only even powers of $\hat{x}$ and $\hat{p}$, it is invariant under $\hat{x}\rightarrow-\hat{x}$ and $\hat{p}\rightarrow-\hat{p}$, and therefore commutes with the parity operator $\hat{\Pi}=e^{i\pi\hat{a}^{\dagger}\hat{a}}$. Thus, $[\hat{H}(t),\hat{\Pi}]=0$ at all times.

For the initial state $\ket{0}$, which has even parity, the dynamics consequently remain restricted to the even Fock sector $\ket{0},\ket{2},\ket{4},\ldots$, while the odd states remain unpopulated. The quartic interaction therefore enlarges the set of allowed transitions within the even-parity manifold without breaking parity conservation.


Substituting the Fock-space expansion $\ket{\psi(t)}=\sum_{k=0}^{N-1}c_k(t)\ket{k}$ into the time-dependent Schr\"odinger equation $i\hbar\,\partial_t\ket{\psi(t)}=\hat{H}(t)\ket{\psi(t)}$ and projecting onto the basis state $\bra{n}$ gives $i\hbar\,dc_n(t)/dt=\sum_{k=0}^{N-1}H_{nk}(t)c_k(t)$, where $H_{nk}(t)=\bra{n}\hat{H}(t)\ket{k}$.

Using the Hamiltonian defined in Eq. \ref{eq:H}, the matrix elements are
\begin{equation}
H_{nk}(t)
=
\frac{1}{2m}\bra{n}\hat{p}^2\ket{k}
+
\frac{m\Omega^2}{8}f(t)\bra{n}\hat{x}^2\ket{k}
+
\frac{\lambda m\Omega^2}{8}f(t)\bra{n}\hat{x}^4\ket{k}.
\label{eq:Hnk}
\end{equation}

Defining the column vector of Fock amplitudes as $\mathbf{c}(t)=(c_0(t),c_1(t),\ldots,c_{N-1}(t))^T$ and the Hamiltonian matrix as $\mathbf{H}(t)=[H_{nk}(t)]_{n,k=0}^{N-1}$, the coupled equations take the compact form
\begin{equation}
i\hbar
\frac{d\mathbf{c}(t)}{dt}
=
\mathbf{H}(t)\mathbf{c}(t).
\label{eq:TDSE_matrix}
\end{equation}

In the natural units as discussed in Sec. \ref{Formulation_QH}, this reduces to $i\,d\mathbf{c}(t)/dt=\mathbf{H}(t)\mathbf{c}(t)$. This finite-dimensional time-dependent Schr\"odinger equation is the central dynamical equation propagated numerically in the following section.

\begingroup
\section{Numerical methods} \label{numerical_methods}
 
\subsection{Fock-space dynamics and number statistics}
\label{sec:num_fock}
 
Throughout the numerical calculations we adopt the natural-unit convention
$m = \hbar = \Omega = 1$, so that $\ell_\Omega = 1$ and the dimensionless time
$\tau = \Omega t$ coincides numerically with $t$; the rf period is then
$T_{\mathrm{rf}} = 2\pi$. The Mathieu parameters are fixed at $a = 0.1$ and
$q = 0.5$ (unless otherwise stated), while the octupolar anharmonicity parameter is varied in the range
$0 \le \lambda \le 5\times10^{-4}$.
 
The matrix form of the Schr\"odinger equation, Eq. \ref{eq:TDSE_matrix}, 
constitutes a system of $N$ coupled first-order ordinary differential equations
for the Fock amplitudes $c_n(\tau)$, where $N$ is the dimension of the truncated
basis. We integrate this system with the explicit adaptive Runge--Kutta (RK45)
method implemented in \texttt{scipy.integrate.solve\_ivp} \cite{virtanen2020scipy},
with absolute and relative error tolerances of $10^{-10}$. The basis size $N$,
the propagation interval, and the number of output points are chosen for each
calculation and are stated with the corresponding results. Short-time
observables are resolved within individual rf periods, whereas the long-time
phase-space and squeezing analysis is carried out over many rf cycles and
evaluated during the 1000th cycle,
$999\,T_{\mathrm{rf}} \le \tau \le 1000\,T_{\mathrm{rf}}$. Quantities at
definite rf phases are sampled at $\tau = jT_{\mathrm{rf}} + \delta\tau$ with
$0 \le \delta\tau \le T_{\mathrm{rf}}$; the phases $0$, $T_{\mathrm{rf}}/4$,
$T_{\mathrm{rf}}/2$, $3T_{\mathrm{rf}}/4$, and $T_{\mathrm{rf}}$ shown in the
phase-space figures therefore refer to local times within the selected cycle.
 
All simulations start from the vacuum state of the reference harmonic
oscillator, $\lvert\psi(0)\rangle = \lvert 0\rangle$, i.e., $c_0(0) = 1$ and
$c_n(0) = 0$ for $n \ge 1$. The state $\lvert 0\rangle$ is the ground state of
the reference oscillator of frequency $\omega_0$ that defines the Fock basis; it
is in general neither the instantaneous ground state of
$\hat{\mathcal{H}}(0)$ nor the ground state of the rf pseudopotential. This
choice ensures that any motional excitation, squeezing, and redistribution among
Fock states arises entirely from the rf drive and the anharmonic perturbation.
 
From the propagated amplitudes we obtain the instantaneous Fock-state
populations $P_n(\tau) = \lvert c_n(\tau)\rvert^2$, from which the mean phonon
number and the phonon-number variance follow as
\begin{equation}
  \begin{aligned}
    \langle n(\tau)\rangle &= \sum_{n=0}^{N-1} n\,P_n(\tau),\\
    \Delta n^2(\tau) &= \langle n^2(\tau)\rangle - \langle n(\tau)\rangle^2,
  \end{aligned}
  \label{eq:n_var}
\end{equation}
with $\langle n^2(\tau)\rangle = \sum_{n=0}^{N-1} n^2 P_n(\tau)$. To compare
with standard number statistics, we use the instantaneous Mandel
parameter \cite{mandel1979subpoissonian}, defined for $\langle n(\tau)\rangle > 0$ as
\begin{equation}
  Q(\tau) = \frac{\Delta n^2(\tau) - \langle n(\tau)\rangle}
                 {\langle n(\tau)\rangle}.
  \label{eq:Mandel}
\end{equation}
The values $Q = 0$, $Q < 0$, and $Q > 0$ correspond to Poissonian,
sub-Poissonian, and super-Poissonian statistics, respectively. Because $Q$ can
become large when $\langle n(\tau)\rangle$ is small, it is always interpreted
together with the corresponding mean phonon number. As reference relations for
instantaneous states, coherent, thermal, and ideal squeezed-vacuum states
satisfy $\Delta n^2 = \langle n\rangle$,
$\Delta n^2 = \langle n\rangle(\langle n\rangle + 1)$, and
$\Delta n^2 = 2\langle n\rangle(\langle n\rangle + 1)$,
respectively \cite{scully1997quantum, mandel1995optical, agarwal2013quantum}.
 
To characterize number-state nonclassicality directly from the Fock
probabilities, we further employ the finite Klyshko witness \cite{klyshko1996observable}
\begin{equation}
  B_n(\tau) = (n+1)\,P_{n-1}(\tau)\,P_{n+1}(\tau) - n\,P_n^2(\tau),
  \quad n \ge 1.
  \label{eq:Klyshko}
\end{equation}
Classical number statistics require $B_n \ge 0$, so that $B_n < 0$ provides a
sufficient signature of nonclassicality. The witness is evaluated from the
instantaneous populations $P_n(\tau)$ at the times $\tau$ indicated in the
corresponding figure.
 
Probability conservation is monitored through the normalization error
$\epsilon_{\mathrm{norm}}(\tau) = \lvert 1 - \lVert \mathbf{c}(\tau)\rVert^2\rvert$,
with $\lVert \mathbf{c}(\tau)\rVert^2 = \sum_{n=0}^{N-1}\lvert c_n(\tau)\rvert^2$,
and the population reaching the edge of the truncated basis through
$P_{\mathrm{top}}(\tau) = \sum_{n=N-5}^{N-1}\lvert c_n(\tau)\rvert^2$. A small
$P_{\mathrm{top}}$ is necessary but not by itself sufficient for convergence,
particularly for observables such as $\Delta n^2$ that weight high-lying Fock
states more strongly. Representative calculations are therefore repeated with
enlarged bases and compared for $\langle n(\tau)\rangle$, $\Delta n^2(\tau)$,
$S_{\min}(\tau)$, and the relevant Fock-state populations. The long-time
phase-space calculations employ $N = 80$; for the population distributions, the
results obtained with $N = 80$ were compared with $N = 100$ and $N = 120$ and
remain unchanged within numerical accuracy, including the high-lying Fock-state
populations shown.

Finally, to separate the rapid micromotion within an rf period from the
response accumulated over many periods, the observables are additionally
averaged in time. \textcolor{black}{Phase-resolved observables during the 1000th RF cycle are evaluated at the specified instantaneous RF phases, whereas the population distributions and the $\lambda$-dependent statistical quantities are constructed from cumulative averages over the first $J$ RF cycles, as specified in the corresponding figures.}
 
\subsection{Quadrature squeezing and phase-space distributions}
\label{sec:num_squeezing}
 
To characterize the motional quadrature fluctuations, we use the generalized
quadrature operator
$\hat{X}_\theta = (\hat{a}\,e^{-i\theta} + \hat{a}^\dagger e^{i\theta})/\sqrt{2}$ \cite{scully1997quantum},
where $\theta$ specifies the quadrature angle in phase space, such that
$\hat{X} = \hat{X}_0$ and $\hat{P} = \hat{X}_{\pi/2}$ are the canonical
quadratures and the vacuum variance is
$(\Delta X_\theta)^2_{\mathrm{vac}} = 1/2$. Since the
Hamiltonian (Eq.~\ref{eq:Hdimless}) is even in $\hat{X}$, parity is conserved, and
for the even-parity initial state $\langle\hat{a}\rangle = 0$, so that
$\langle\hat{X}_\theta\rangle = 0$. Following the standard
quadrature-squeezing criterion, the squeezing parameter then takes the form
\begin{equation}
  S_\theta(\tau) = \left[\Delta X_\theta(\tau)\right]^2 - \frac{1}{2}
  = \langle n(\tau)\rangle
  + \mathrm{Re}\!\left[e^{-2i\theta}\langle\hat{a}^2(\tau)\rangle\right],
  \label{eq:Stheta}
\end{equation}
where the anomalous moment is evaluated directly from the Fock amplitudes as
$\langle\hat{a}^2(\tau)\rangle = \sum_{n=0}^{N-3}\sqrt{(n+1)(n+2)}\,
c_n^*(\tau)\,c_{n+2}(\tau)$. The full angular dependence of the quadrature
fluctuations is thus obtained without reconstructing the spatial wave function.
The state is quadrature squeezed whenever $S_\theta < 0$ for at least one
$\theta$, while $S_\theta = 0$ corresponds to the vacuum level and
$S_\theta \ge -1/2$ follows from the non-negativity of the
variance \cite{strandberg2021wigner}.
 
The strongest squeezing at each time is quantified by
$S_{\min}(\tau) = \min_\theta S_\theta(\tau)$ \cite{strandberg2021wigner}. Writing
$\langle\hat{a}^2\rangle = \lvert\langle\hat{a}^2\rangle\rvert e^{i\phi}$,
Eq.~\eqref{eq:Stheta} becomes
$S_\theta = \langle n\rangle + \lvert\langle\hat{a}^2\rangle\rvert
\cos(\phi - 2\theta)$, and the minimum follows analytically as
\begin{equation}
  S_{\min}(\tau) = \langle n(\tau)\rangle
  - \lvert\langle\hat{a}^2(\tau)\rangle\rvert,
  \label{eq:Smin}
\end{equation}
attained at the optimal squeezing angle
$\theta_s = [\arg\langle\hat{a}^2\rangle - \pi]/2 \pmod{\pi}$. The orthogonal
anti-squeezed quadrature lies at $\theta_{as} = \theta_s + \pi/2 \pmod{\pi}$,
with $S_{\max} = \langle n\rangle + \lvert\langle\hat{a}^2\rangle\rvert$, so
that the minimum and maximum quadrature variances are
$(\Delta X_s)^2 = S_{\min} + 1/2$ and $(\Delta X_{as})^2 = S_{\max} + 1/2$. In
the numerical calculations, $S_\theta$ is evaluated over $0 \le \theta < \pi$,
which contains all physically distinct quadrature directions; the full curve is
used for visualization, whereas the reported $S_{\min}$ and $\theta_s$ are
obtained from the analytical expressions above. Angles in the figures are given
in degrees. For the long-time analysis, $S_\theta$, $S_{\min}$, and $\theta_s$
are evaluated at the local phases $0$, $T_{\mathrm{rf}}/4$,
$T_{\mathrm{rf}}/2$, $3T_{\mathrm{rf}}/4$, and $T_{\mathrm{rf}}$ within the
1000th RF cycle.
 
For selected times, the evolved pure state
$\hat{\rho}(\tau) = \lvert\psi(\tau)\rangle\langle\psi(\tau)\rvert$ is
represented by its Wigner quasiprobability distribution $W(X,P)$, with the
principal squeezing and anti-squeezing directions given by $\theta_s$ and
$\theta_{as}$. The probability distribution of a rotated quadrature is obtained
from the corresponding marginal of the Wigner function,
$P(X_\theta) = \int W(X_\theta, X_{\theta+\pi/2})\,dX_{\theta+\pi/2}$ \cite{wigner1932quantum},
and is compared with the vacuum distribution
$P_{\mathrm{vac}}(X_\theta) = \pi^{-1/2}e^{-X_\theta^2}$ of variance $1/2$. A
distribution narrower than this reference along the optimal squeezing
direction is therefore consistent with $S_{\min} < 0$.

\begingroup
\section{RESULTS AND DISCUSSION} \label{results_and_discussions}

We now examine how the octupole-induced quartic contribution modifies the quantum dynamics of the trapped ion. \textcolor{black}{The results are discussed sequentially in terms of the Fock-state distribution, number-resolved nonclassicality, phonon-number statistics, and phase-space and quadrature dynamics.} The quadratic RF trap provides the reference periodically driven oscillator, whereas the quartic term introduces both additional couplings between Fock states and an occupation-dependent nonlinear phase evolution.

Figure~\ref{fig:population_dist} shows the cumulative time-averaged populations $\overline{P_n}$ of selected even Fock states. The horizontal axis gives the Fock number $n$, while the vertical axis shows the corresponding population on a logarithmic scale. The upper panels represent the quadratic RF trap and the lower panels the RF-modulated quartic trap with $\lambda=5\times10^{-4}$. From left to right, the averaging interval extends over the first $100$, $500$, and $1000$ RF cycles.
\begin{figure}[htbp]
    \centering
    \includegraphics[width=1\linewidth]{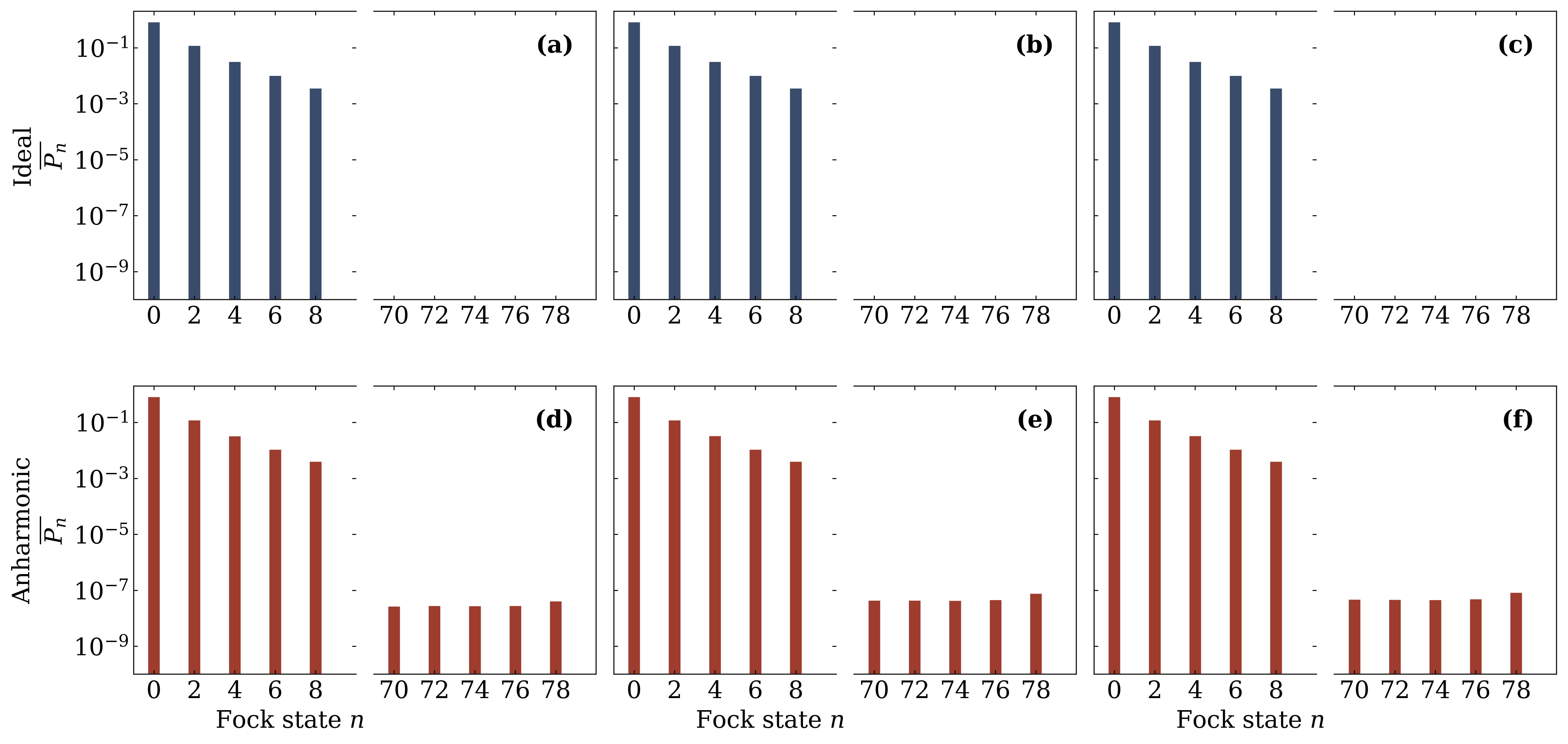}
    \caption{Time-averaged even-Fock-state population $\overline{P_n}$ for the trapped ion under different trapping conditions and evolution intervals with $a=0.1$, $q=0.5$, and $N=80$ with cumulative averages over RF cycles $1$--$100$ [(a) and (d)], $1$--$500$ [(b)and (e)], and $1$--$1000$ [(c) and (f)]. Panels \textbf{(a)--(c)} correspond to the driven quadratic trap ($\lambda = 0$). Panels \textbf{(d)--(f)} show the corresponding results in the presence of the RF-modulated quartic contribution ($\lambda=5\times10^{-4}$). Selected low-lying even states $n=0,2,4,6,8$ and high-lying states $n=70$--$78$ are displayed.}
    \label{fig:population_dist}
\end{figure}
In both cases the population is concentrated predominantly in the lowest even Fock states, while the odd states remain unoccupied. This follows from parity conservation, since the Hamiltonian contains only even powers of $\hat x$ and $\hat p$. The important difference is the structure of the high-$n$ tail. In the quadratic trap the population decreases extremely rapidly with $n$, whereas the quartic trap develops a considerably broader, although still weak, population tail extending to high even states. 

The origin of this difference is richer than the additional $\Delta n=\pm4$ selection rule alone. The quadratic interaction connects states through $\Delta n=\pm2$, whereas the quartic operator contains $\Delta n=\pm2$ and $\Delta n=\pm4$ matrix elements. Moreover, these quartic matrix elements increase rapidly with $n$; asymptotically both the $\Delta n=2$ and $\Delta n=4$ couplings scale approximately as $n^2$. Thus, once a small amplitude reaches higher states, the nonlinear interaction can couple that part of the wave function increasingly strongly as discussed in Appendix \ref{mat_ele} . At the same time, the diagonal quartic matrix element,
$$
\langle n|\hat x^4|n\rangle
\propto
6n^2+6n+3,
$$
produces an $n$-dependent diagonal contribution. The anharmonic dynamics therefore results from a competition between enhanced Fock-space connectivity and nonlinear detuning of the participating states. 
The high-$n$ tail (Fig.~\ref{fig:population_dist}~(d)--(f)) reflects coherent RF-driven excitation and redistribution within the reference-oscillator basis, rather than dissipative heating. Also note, the cumulative time-averaged distribution is essentially independent of the averaging window confirming that motional state reaches a stationary quasi-periodic distribution.
\begin{figure}[htbp]
    \centering
    \includegraphics[width=0.98\linewidth]{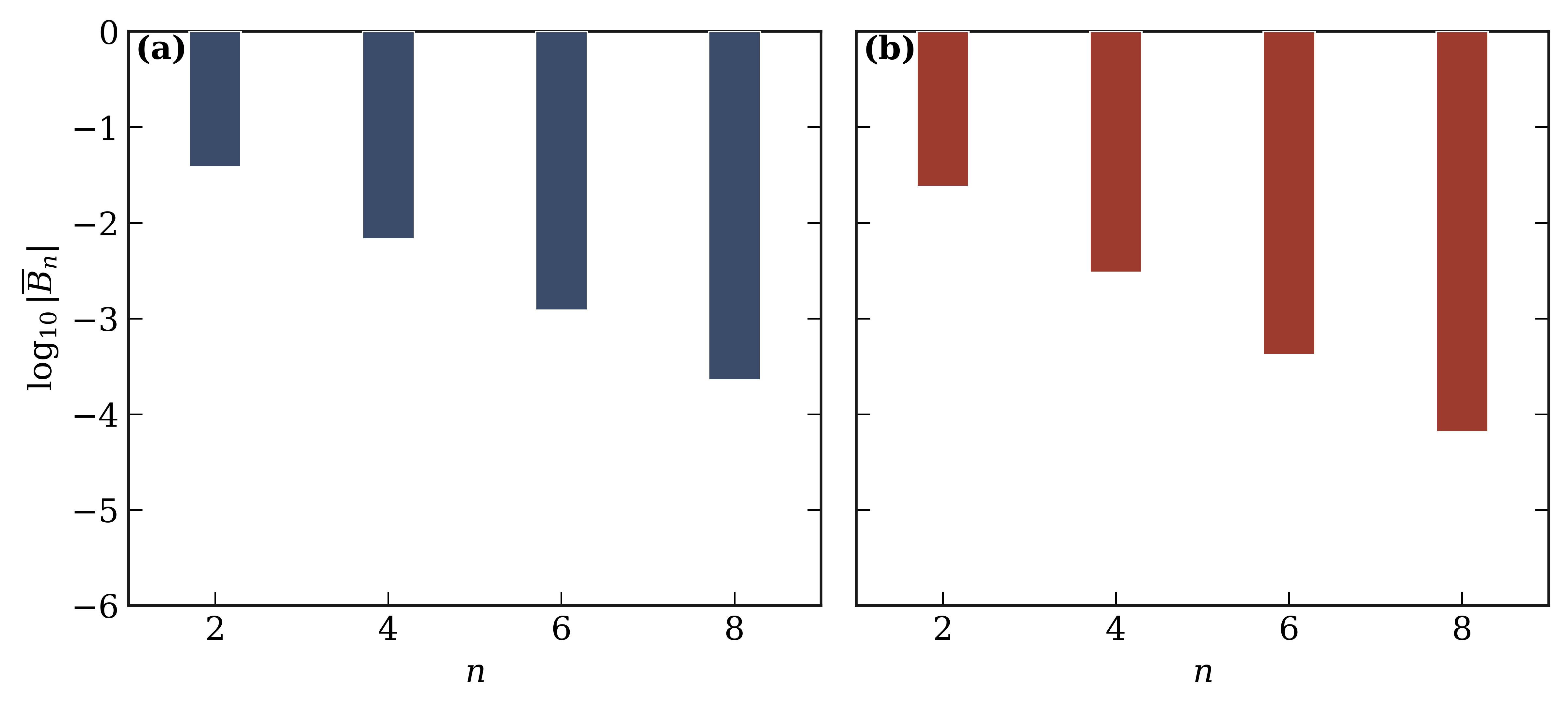}
    \caption{Logarithmic magnitude $\log_{10}|B_n|$ of the finite Klyshko witness for selected even Fock states after the long-time evolution. \textbf{(a)} corresponds to the driven quadratic trap, while \textbf{(b)} corresponds to the RF-modulated quartic trap. The calculations use $a=0.1$, $q=0.5$, $N=80$, and $\lambda=5\times10^{-4}$ for the anharmonic case.}
    \label{fig:klyshko}
\end{figure}
The populations shown near the upper end of the $N=80$ basis were checked by repeating the calculation with $N=100$ and $N=120$, for which the displayed distribution remains unchanged within numerical accuracy. The high-$n$ tail in Fig.~\ref{fig:population_dist}~((d)--(f)) is therefore not a truncation artifact.

To examine whether the populated Fock states identified in Fig.~\ref{fig:population_dist} exhibit nonclassical number statistics, Fig.~\ref{fig:klyshko} shows the logarithmic magnitude $\log_{10}|B_n|$ of the Klyshko witness (see Eq.~\ref{eq:Klyshko}) for a few selected even Fock states. Fig.~\ref{fig:klyshko}(a) corresponds to the quadratic RF trap, whereas Fig.~\ref{fig:klyshko}(b) shows the RF-modulated quartic trap. In both trapping regimes the displayed populated even states satisfy $B_n<0$, demonstrating violation of the Klyshko classicality inequality and indicating nonclassical number statistics.

This result follows directly from the parity structure established above. Since the dynamics preserve even parity, $P_{2k+1}=0$, and for any populated even state with $n>0$ the Klyshko witness (Eq. \ref{eq:Klyshko}) reduces to $B_n=-nP_n^2<0$. Thus, the important information conveyed by Fig.~\ref{fig:klyshko} is that the populated motional states remain nonclassical in both the ideal quadratic and RF-modulated anharmonic regimes. Since $|B_n|=nP_n^2$ in the present parity-conserving dynamics, its magnitude should not be interpreted as a quantitative measure of the degree of nonclassicality.

Although these high-lying populations are individually small, they can have a disproportionately large influence on higher moments of the phonon distribution. To quantify this effect, Fig.~\ref{fig:lambda_mean_variance} shows the mean reference-oscillator occupation $\overline{\langle n\rangle}$ and the corresponding variance $\overline{\Delta n^2}$ as functions of $\lambda$. Both quantities are averaged over the first $1000$ RF cycles.
\begin{figure}[htbp]
    \centering
    \includegraphics[width=0.72\linewidth]{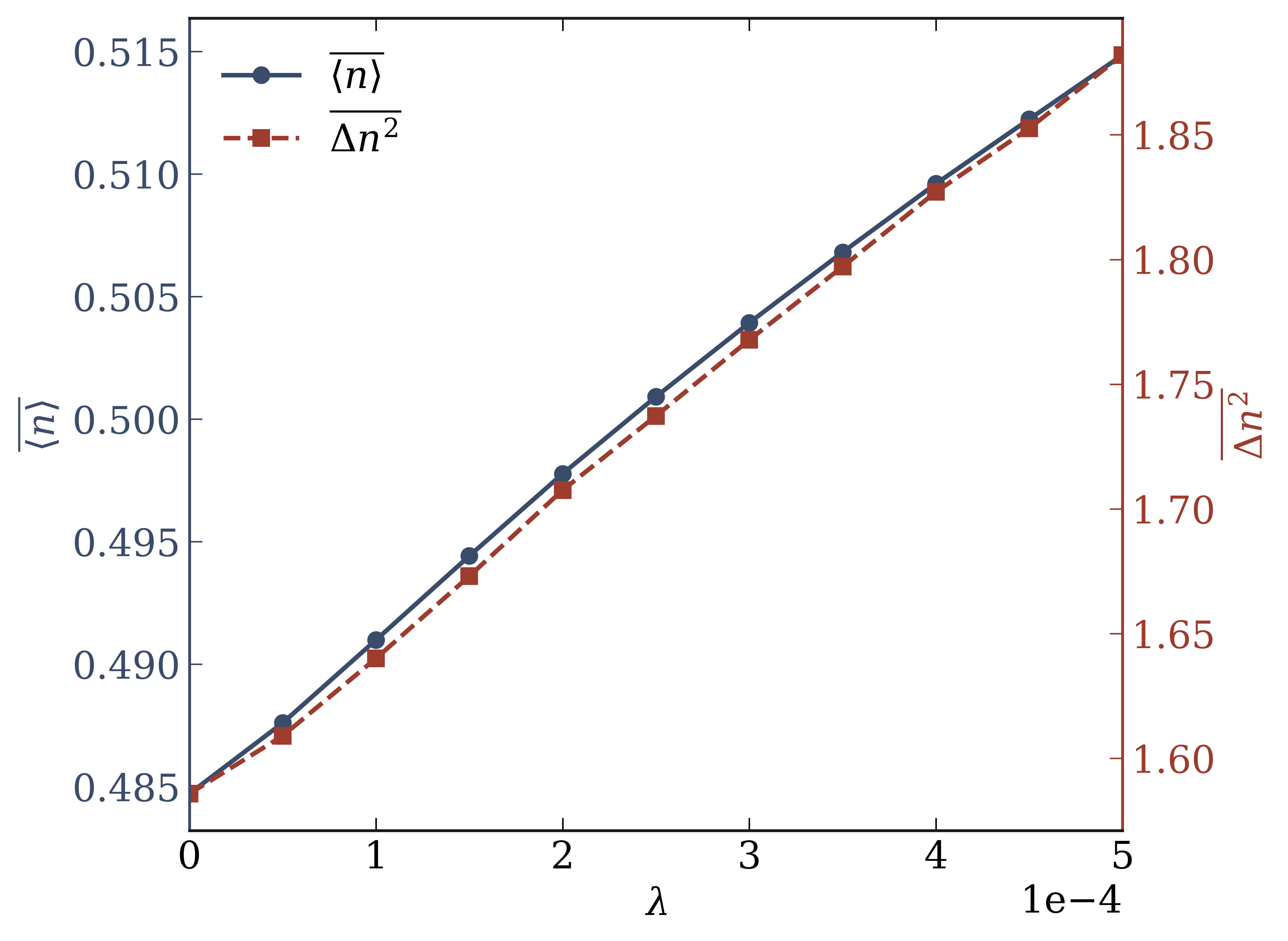}
    \caption{Mean phonon number $\overline{\langle n\rangle}$ and phonon-number variance $\overline{\Delta n^2}$, averaged over the first $1000$ RF cycles, as functions of the dimensionless quartic coupling $\lambda$ for $a=0.1$ and $q=0.5$.}
    \label{fig:lambda_mean_variance}
\end{figure}
Both the mean and the variance increase monotonically with the quartic coupling. The mean changes from approximately $\overline{\langle n\rangle}\simeq0.485$ at $\lambda=0$ to $\overline{\langle n\rangle}\simeq0.515$ at $\lambda=5\times10^{-4}$, whereas the variance increases from approximately $\overline{\Delta n^2}\simeq1.59$ to $\overline{\Delta n^2}\simeq1.88.$ Thus, the mean changes by only about $6\%$, while the variance increases by about $18\%$. The anharmonicity therefore modifies the width of the number distribution substantially more strongly than its center.

This difference follows naturally from the Fock-space redistribution in Fig.~\ref{fig:population_dist}. Contributions to the mean are weighted by $nP_n$, whereas those entering $\langle n^2\rangle$ are weighted by $n^2P_n$. Consequently, a weak population transferred to high $n$ has a relatively small effect on $\overline{\langle n\rangle}$ but a much stronger effect on $\overline{\Delta n^2}$. The variance is therefore a sensitive probe of the weak anharmonic tail.

The diagonal quartic term provides a second contribution to this behavior. Since the diagonal anharmonic contribution varies approximately as $n^2$, the difference between the corresponding diagonal contributions for states separated by two quanta becomes occupation dependent. In particular,
$$
\delta H^{(4)}_{n+2,n+2}(t)
-
\delta H^{(4)}_{n,n}(t)
\propto
24n+36.
$$

The effective detuning of the RF-induced transitions therefore changes as the state spreads through Fock space. This is the quantum analogue of the amplitude-dependent oscillation frequency of a classical Duffing-type oscillator. The octupole field consequently changes not only which states are coupled but also the resonance conditions encountered during the evolution.

To display directly how the two statistical moments evolve together, Fig.~\ref{fig:variance_mean_lambda} plots $\overline{\Delta n^2}$ as a function of $\overline{\langle n\rangle}$, with the corresponding value of $\lambda$ indicated by the color scale. The points trace a smooth monotonic path toward larger mean occupation and larger fluctuations as $\lambda$ is increased.
\begin{figure}[htbp]
    \centering
    \includegraphics[width=0.62\linewidth]{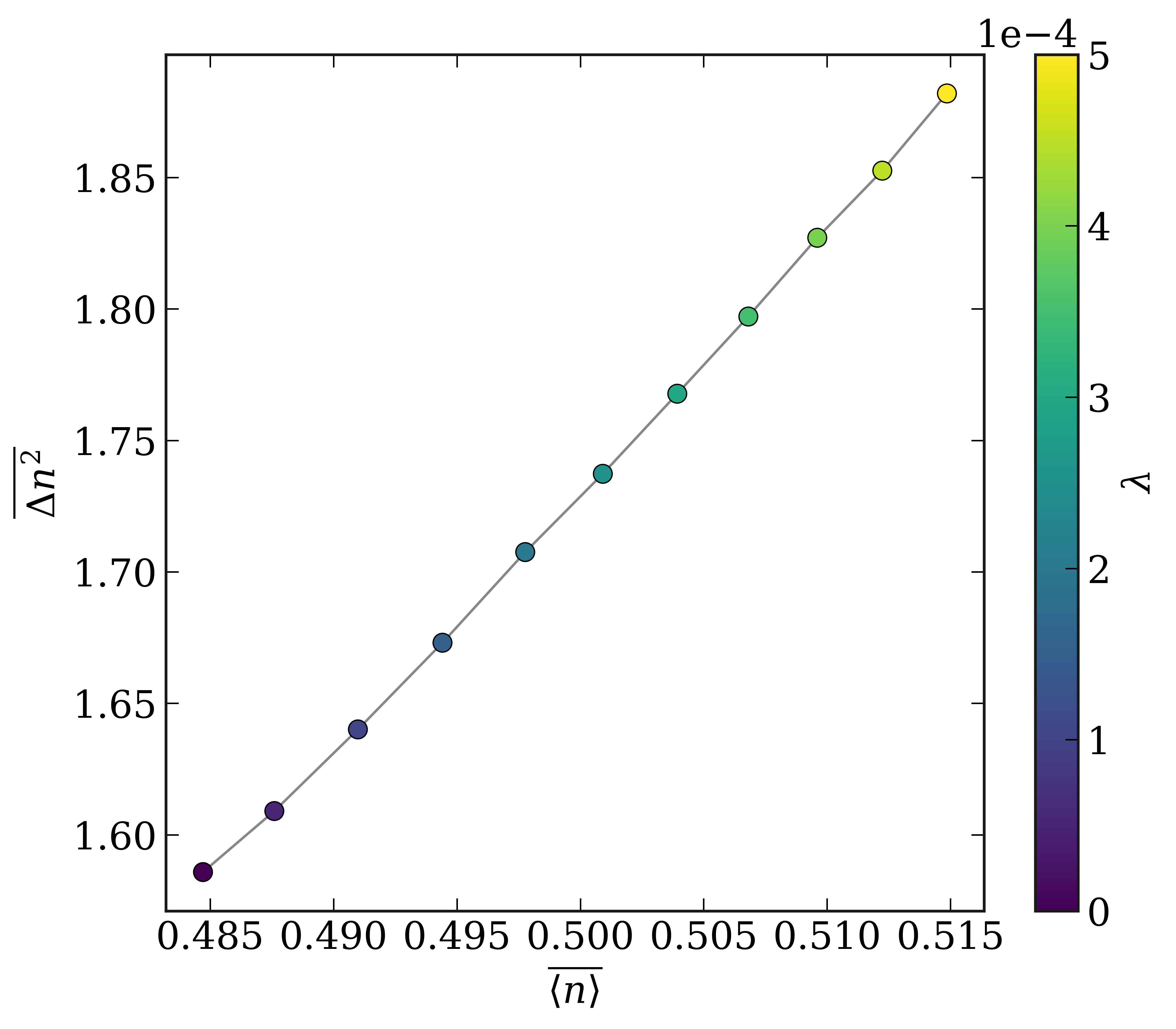}
    \caption{Phonon-number variance $\overline{\Delta n^2}$ as a function of the mean phonon number $\overline{\langle n\rangle}$, with both quantities averaged over the first $1000$ RF cycles, as $\lambda$ is varied from $0$ to $5\times10^{-4}$. The color scale indicates the corresponding value of $\lambda$.}
    \label{fig:variance_mean_lambda}
\end{figure}
The nearly linear appearance of this trajectory indicates that the system remains in a weak-anharmonicity regime over the parameter interval considered here. The quartic perturbation continuously deforms the number distribution rather than producing an abrupt change of dynamical behavior. The slope of the trajectory nevertheless emphasizes that the variance is more responsive than the mean, consistent with the progressive development of the weak high-$n$ tail. This approximately linear relation is specific to the restricted range studied here and is not expected to remain universal at stronger anharmonicity, where nonlinear detuning becomes more important.

The variance is already significantly larger than the mean at $\lambda=0$. This reflects the fact that the quadratic RF trap does not generate coherent-state statistics. The time-dependent quadratic confinement acts parametrically and produces correlated pairs of oscillator excitations. For an instantaneous zero-displacement squeezed vacuum,
$$
\Delta n^2
=
2\langle n\rangle
\left(
\langle n\rangle+1
\right),
$$

so large super-Poissonian fluctuations are expected even without the quartic perturbation. In the present calculations the plotted quantities are long-time averages, and the RF micromotion further contributes to the difference between the averaged moments and the corresponding instantaneous squeezed-vacuum relation as discussed in Appendix \ref{Sq_vacc}.

The relative strength of these number fluctuations is summarized in Fig.~\ref{fig:mandel_lambda}, which shows the cycle averaged Mandel $Q_{\rm}$ parameter (Eq. \ref{eq:Mandel}), where $Q = \frac{\overline{\Delta n^2}-\overline{\langle n\rangle}
}{\overline{\langle n\rangle}}$,
as a function of $\lambda$. The value remains positive throughout the investigated interval and increases from approximately $2.27$ to $2.65$ as the quartic coupling is increased.
\begin{figure}[htbp]
    \centering
    \includegraphics[width=0.62\linewidth]{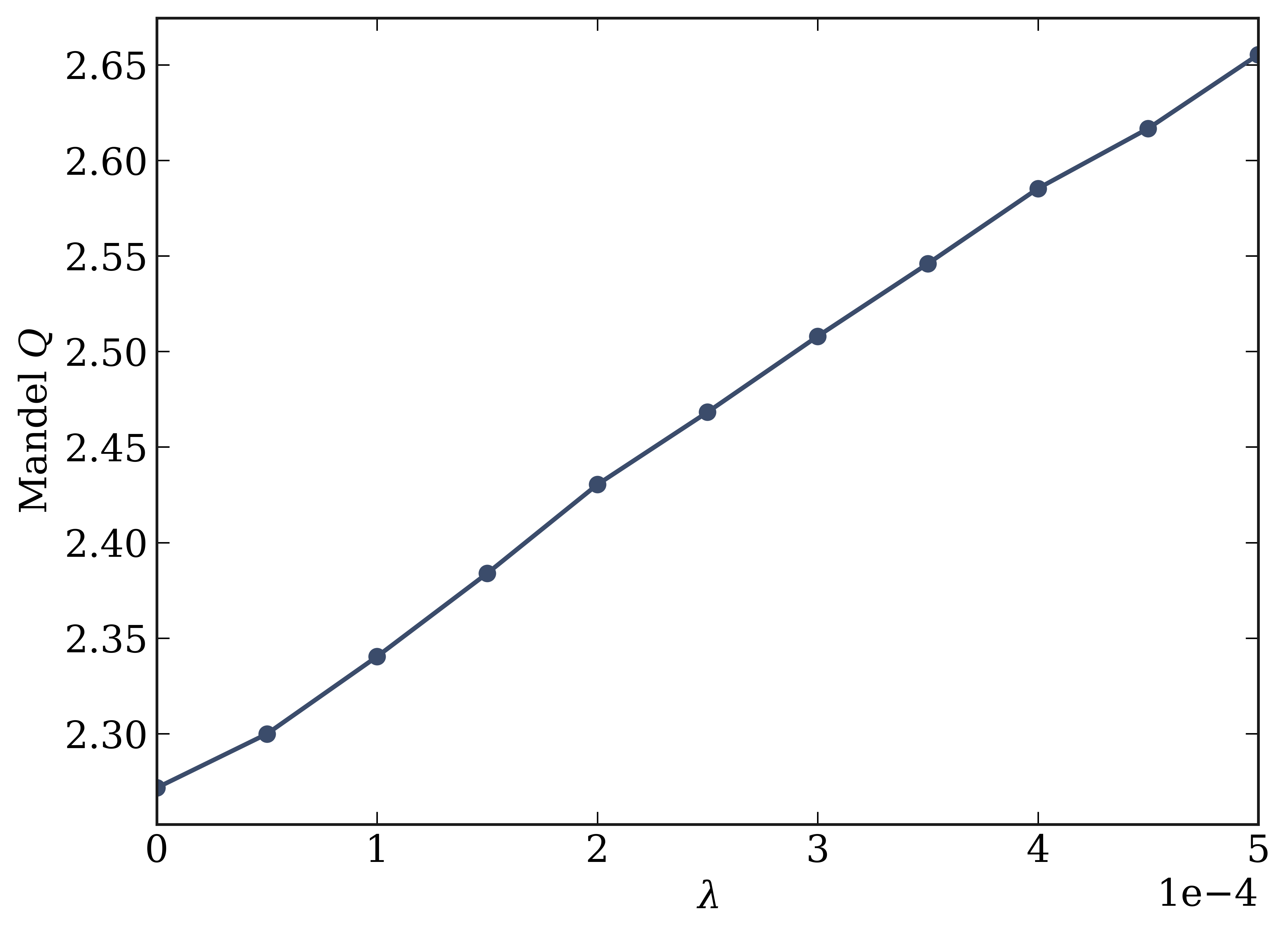}
    \caption{Cycle averaged Mandel $Q_{\rm}$ parameter constructed from the phonon-number moments averaged over the first $1000$ RF cycles, as a function of the quartic coupling $\lambda$ for $a=0.1$ and $q=0.5$.}
    \label{fig:mandel_lambda}
\end{figure}
\textcolor{black}{The positive value of $Q$ at $\lambda=0$ is consistent with the pair-correlated squeezed dynamics of the quadratic RF oscillator, rather than being a consequence of anharmonicity.} The increase with $\lambda$ instead measures the additional broadening introduced by the quartic interaction. \textcolor{black}{Thus Figs.~\ref{fig:lambda_mean_variance}--\ref{fig:mandel_lambda} consistently show that weak octupole anharmonicity broadens the population distribution within the even Fock ladder, with a stronger effect on the number fluctuations than on the mean phonon occupation.}

A positive Mandel parameter alone is not a sufficient criterion for nonclassicality, since classical mixtures may also be super-Poissonian. We therefore next examine the phase-sensitive quantum fluctuations of the motional state.

\begin{figure*}[p]
    \centering
    \includegraphics[width=0.95\linewidth]{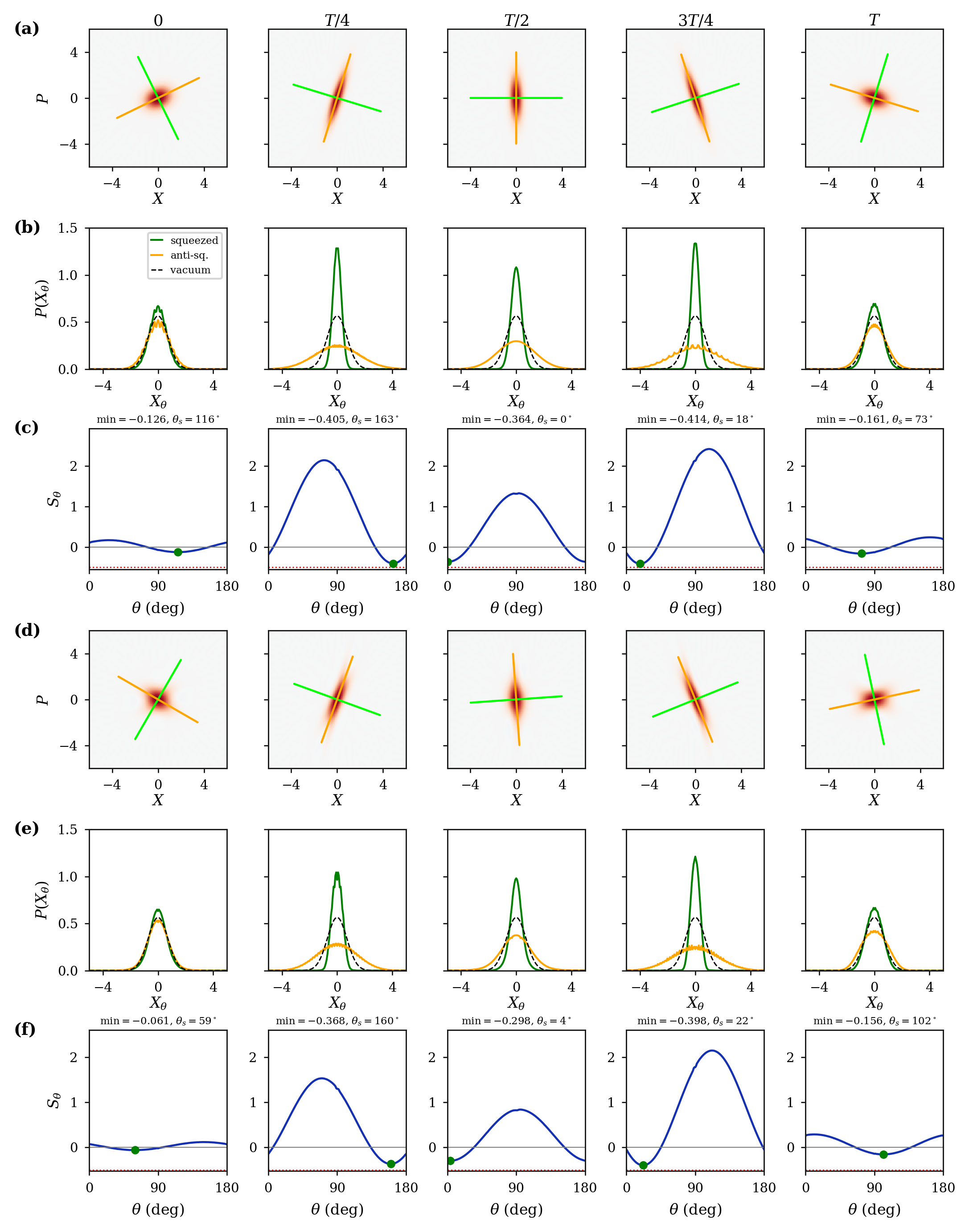}
    \caption{Phase-space and quadrature dynamics during the $1000$th RF cycle for the quadratic reference trap with $\lambda=0$~(a--c) and the RF-modulated quartic trap with $\lambda=5\times10^{-4}$~(d--f). (a) and (d): Wigner distributions; (b) and (e): corresponding squeezed and anti-squeezed quadrature marginals; (c) and (f): angular dependence of $S_\theta$.}
    \label{fig:wigner_combined}
\end{figure*}

Figure~\ref{fig:wigner_combined} compares the phase-space and quadrature dynamics of the quadratic reference trap [panels (a) -- (c)] and the RF-modulated quartic trap [panels (c) -- (f)] during the $1000$th RF cycle with $a=0.1$, $q=0.5$, $\omega_0=\sqrt{a}/2$, and $N=80$. For each trapping configuration, the five columns correspond to the local RF phases $0$, $T_{\rm rf}/4$, $T_{\rm rf}/2$, $3T_{\rm rf}/4$, and $T_{\rm rf}$. Panels (a) and (d) show the Wigner distributions together with the principal squeezed and anti-squeezed directions, whereas panels (b) and (e) illustrate the squeezed and anti-squeezed quadrature marginals, $P(X_\theta)$, together with the vacuum reference. The extraction of the rotated-quadrature marginal from the Wigner function is discussed in Appendix~\ref{app:marginal}. Panels (c) and (f) show the squeezing parameter $S_\theta=(\Delta X_\theta)^2-\tfrac{1}{2}$ as a function of the quadrature angle $\theta$, with the marker indicating $\min[S_\theta]$ at the optimal squeezing angle $\theta_s$. 

For the quadratic trap, the Wigner distribution remains predominantly elliptical while its eccentricity and principal axes evolve during the RF cycle. The quadrature marginals show the corresponding compression and broadening along the principal directions, and the minima of $S_\theta$ remain below zero at all five displayed phases. This behavior is characteristic of the Gaussian dynamics generated by a time-dependent quadratic Hamiltonian. In this case the Heisenberg evolution has the Bogoliubov form \cite{lewis1969exact} $\hat a(t)=u(t)\hat a(0)+v(t)\hat a^\dagger(0)$, so that the RF modulation generates phase-sensitive pair correlations while preserving the Gaussian character of the initial vacuum state. The magnitude of $\langle\hat a^2\rangle$ determines the strength of the squeezing, whereas its phase determines the orientation of the squeezing ellipse.

Panels (c)--(f) show that the sub-vacuum quadrature fluctuations persist when the RF-modulated quartic interaction is included. The Wigner distributions continue to rotate and deform through the RF cycle, while the locations and depths of the minima of $S_\theta$ and the corresponding optimal quadrature directions are modified relative to the quadratic reference case. The quartic interaction therefore changes the detailed covariance and phase-space structure without destroying the RF-generated squeezing for the coupling considered here.

The physical origin of these modifications lies in the nonlinear structure of the quartic interaction. Its off-diagonal matrix elements redistribute amplitude through the even Fock ladder, while the diagonal contribution produces occupation-dependent phase accumulation. Since $\langle\hat a^2\rangle=\sum_n\sqrt{(n+1)(n+2)}\,c_n^*c_{n+2}$ depends on both the amplitudes and relative phases of neighboring even Fock components, the nonlinear evolution alters both the magnitude and orientation of the quadrature squeezing. The quartic term also removes the guarantee of exact Gaussian evolution associated with a purely quadratic Hamiltonian, allowing nonlinear phase-space shearing to modify the Wigner distribution.

Figure~\ref{fig:wigner_combined} therefore establishes that quadrature squeezing remains robust in both trapping configurations while its detailed phase dependence is altered by the octupole contribution. Since this figure compares only $\lambda=0$ and $\lambda=5\times10^{-4}$, however, it does not by itself establish how $S_{\min}$ evolves continuously with anharmonicity. This dependence is examined directly in Fig.~\ref{fig:minS_lambda}.

\begin{figure}[htbp]
    \centering
    \includegraphics[width=0.62\linewidth]{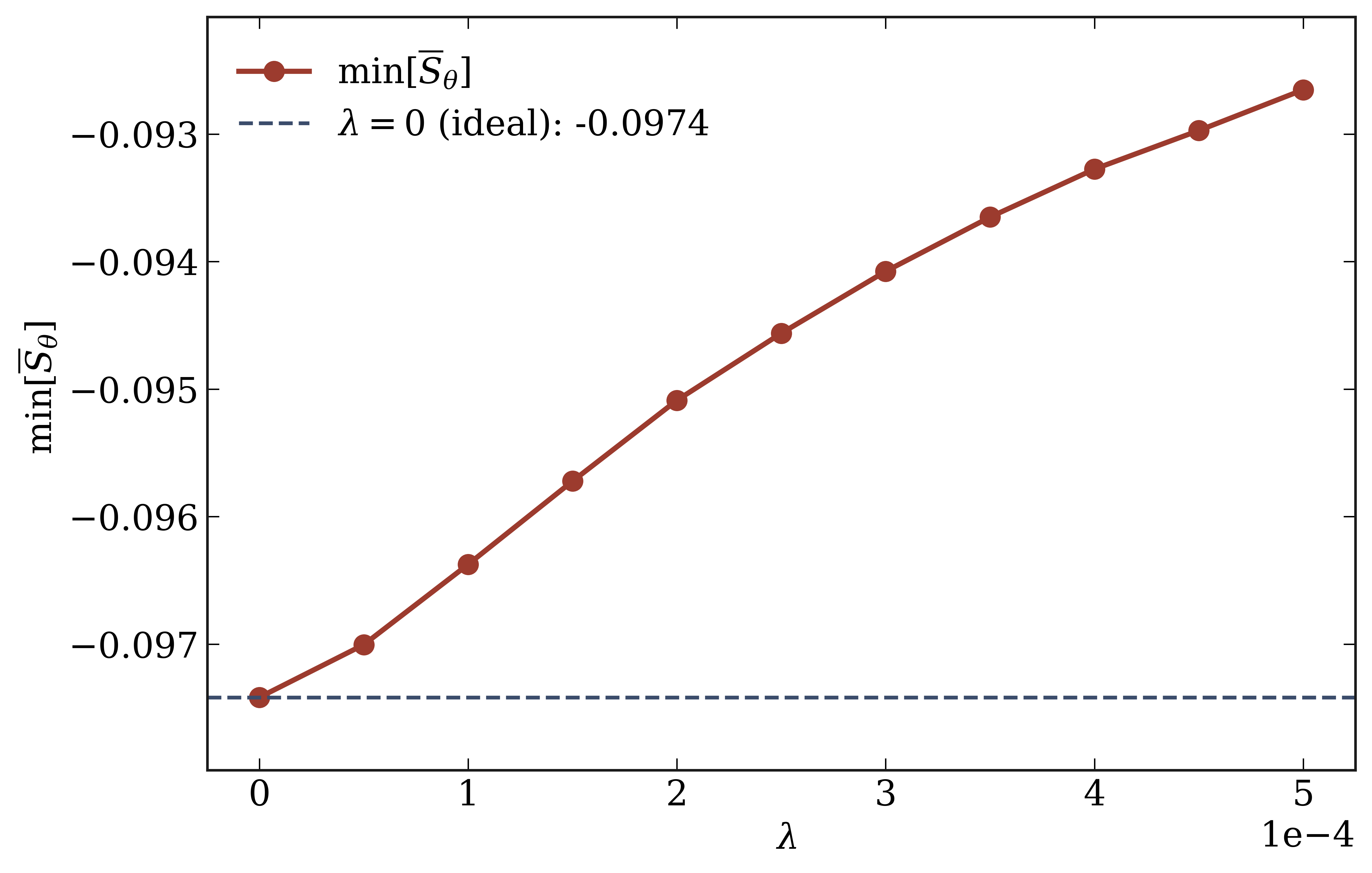}
    \caption{Minimum of the cumulatively averaged quadrature-squeezing parameter,
$\min\!\left[\bar{S_{\theta}}\right]$, evaluated over the first
1000 RF cycles as a function of the dimensionless quartic coupling $\lambda$ for
$a=0.1$ and $q=0.5$. The dashed horizontal line indicates the corresponding
value for the quadratic reference trap, $\lambda=0$.}
    \label{fig:minS_lambda}
\end{figure}

To examine the overall influence of the anharmonicity on the quadrature squeezing over the long-time evolution, Fig.~\ref{fig:minS_lambda} shows the cumulative average $\min\!\left[\bar{S_{\theta}}\right]$ over the first $1000$ RF cycles as a function of the quartic coupling $\lambda$. The dashed horizontal line denotes the corresponding value for the quadratic reference trap at $\lambda=0$.

The averaged minimum squeezing remains negative throughout the investigated range $0\leq\lambda\leq5\times10^{-4}$, demonstrating that sub-vacuum quadrature fluctuations persist over the complete weak-anharmonicity regime considered here. As $\lambda$ is increased, $\min\!\left[\bar{S_{\theta}}\right]$ becomes progressively less negative over the parameter range shown, indicating a modest weakening of the time-averaged squeezing while preserving its nonclassical character.

This behavior complements the phase-resolved dynamics shown in Fig.~\ref{fig:wigner_combined}. Whereas Fig.~\ref{fig:wigner_combined} demonstrates that the instantaneous squeezing strength and orientation vary strongly with the RF phase, Fig.~\ref{fig:minS_lambda} summarizes the net long-time response to the quartic interaction. The gradual change of $\min\!\left[\bar{S_{\theta}}\right]$ is consistent with the occupation-dependent phase evolution introduced by the quartic term, which modifies the coherent neighboring-state correlations entering $\langle\hat a^2\rangle$. Thus, the octupole perturbation alters the accumulated phase-sensitive coherence while the quadrature squeezing remains robust over the full range of anharmonicities studied.

Taken together, the results reveal two complementary consequences of the octupole interaction. Its off-diagonal matrix elements increase the connectivity of the even Fock ladder and broaden the phonon-number distribution, while its diagonal $n^2$ dependence introduces occupation-dependent energy shifts and nonlinear phase dispersion. The former is manifested most clearly in the population tail, the variance, and the Mandel parameter, whereas the latter modifies the phase-sensitive coherence responsible for quadrature squeezing. The RF-driven quadratic trap therefore realizes a periodically squeezed Gaussian oscillator, while the octupole correction converts the dynamics into that of a weakly nonlinear driven oscillator in which Fock-space spreading and phase-space shearing occur simultaneously. In this regime the number fluctuations are enhanced, whereas the quadrature squeezing remains robust; its long-time averaged magnitude is modestly weakened and its phase-resolved orientation is modified by the anharmonic interaction. These two responses provide complementary signatures of weak anharmonicity in the quantum motion of a trapped ion.

\section{Conclusion} \label{Conclusions}

We have investigated the quantum dynamics of a single ion in an RF Paul trap by retaining the explicit time dependence of the quadrupole confinement and incorporating a weak octupole-induced quartic perturbation. Within a Fock-basis formulation, the quadratic RF interaction couples states through $\Delta n=\pm2$, whereas the quartic contribution opens additional $\Delta n=\pm4$ channels while preserving the parity. The octupole interaction consequently redistributes population over a broader range of even Fock states and enhances the phonon-number fluctuations, with the variance showing a stronger response to the anharmonicity than the mean occupation.

The phase-space analysis shows that the quadratic RF trap generates a rotating squeezed Gaussian motional state, while the weak quartic perturbation modifies the detailed Wigner distribution and the magnitude and orientation of the quadrature squeezing. Moreover, the long-time averaged minimum squeezing parameter becomes progressively less negative with increasing $\lambda$, indicating a gradual weakening of the quadrature squeezing as the octupole strength is increased. Importantly, the long-time averaged minimum squeezing remains below the vacuum level throughout the investigated range of $\lambda$, demonstrating that quadrature nonclassicality survives in the presence of the weak octupole field. The Klyshko criterion independently confirms nonclassical number statistics in both the quadratic and anharmonic trapping regimes. Overall, these results show that weak octupole corrections can appreciably modify the population statistics and phase-space dynamics of a trapped ion while preserving the nonclassical character of its motional state, thereby extending the quantum description of RF confinement beyond the ideal quadrupole approximation.

The present analysis is restricted to weak octupolar anharmonicity, for which the required Fock-space basis remains computationally tractable within the numerical framework employed here. An important direction for future work is therefore to extend the analysis to stronger anharmonicities and to systematically investigate the evolution of the motional nonclassicality beyond the weak-perturbation regime. Such an extension would allow one to determine whether the progressive modification of the Fock-state populations, number fluctuations, Wigner distribution, and quadrature squeezing observed here eventually leads to a qualitative change or suppression of nonclassical behavior. Exploring this crossover, together with higher values of the Mathieu parameter $q$, would provide a more complete picture of the interplay between RF-driven confinement, anharmonicity, and nonclassical motional dynamics. These calculations would, however, require substantially larger Fock-space bases and careful convergence analysis. The inclusion of atom-photon interaction, finite decay and other dissipative processes would provide a further step toward describing the dynamics under experimentally realistic conditions.

\section{Acknowledgements}

 SJ and AD acknowledges Indian Association for the Cultivation of Science India for research fellowship. ND thanks SERB/ANRF India (CRG/2023/001529) and BRNS India (58/14/21/2023-BRNS12329) for funding. 

\section{Conflicts of Interest}
There are no conflicts to declare. 

\section{Data Availability Statement}
The data that support the findings of this study are available upon reasonable request.

\appendix 
\section{Matrix Elements} \label{mat_ele}
\label{Matrix Elements}

The non-zero matrix elements of the relevant operators in the Fock basis are computed analytically using the ladder-operator representation. Defining
$$
x_s
\equiv
\frac{\hbar}{2m\omega_0}
=
x_{\rm sc}^2,
$$
the diagonal matrix elements of $\hat{x}^2$ are
$$
\bra{n}\hat{x}^2\ket{n}
=
x_s(2n+1),
$$
while the upper off-diagonal elements are
$$
\bra{n+2}\hat{x}^2\ket{n}
=
x_s
\sqrt{(n+1)(n+2)}.
$$

For the quartic operator, the diagonal elements are
$$
\bra{n}\hat{x}^4\ket{n}
=
x_s^2
\left(
6n^2+6n+3
\right),
$$
the $\Delta n=2$ matrix elements are
$$
\bra{n+2}\hat{x}^4\ket{n}
=
x_s^2
(4n+6)
\sqrt{(n+1)(n+2)},
$$
and the $\Delta n=4$ matrix elements are
$$
\bra{n+4}\hat{x}^4\ket{n}
=
x_s^2
\sqrt{
(n+1)(n+2)(n+3)(n+4)
}.
$$

The corresponding matrix elements of $\hat{p}^2$ are
$$
\bra{n}\hat{p}^2\ket{n}
=
\frac{m\hbar\omega_0}{2}
(2n+1),
$$
and
$$
\bra{n+2}\hat{p}^2\ket{n}
=
-
\frac{m\hbar\omega_0}{2}
\sqrt{(n+1)(n+2)}.
$$

Since $\hat{x}^2$, $\hat{x}^4$, and $\hat{p}^2$ are Hermitian, the corresponding lower off-diagonal matrix elements follow from
$$
\bra{n}\hat{O}\ket{n+r}
=
\bra{n+r}\hat{O}\ket{n}^{*},
$$
where $\hat{O}=\hat{x}^2,\hat{x}^4,\hat{p}^2$ and $r=2$ or $4$ as appropriate. All matrix elements outside these allowed bands vanish identically.

\section{Relation between instantaneous and time-averaged squeezed-vacuum statistics} \label{Sq_vacc}

For a zero-displacement squeezed-vacuum state, the instantaneous phonon-number variance satisfies
\begin{equation}
\Delta n^2(t)
=
2\langle n(t)\rangle
\left[
\langle n(t)\rangle+1
\right].
\label{eq:instant_sqvac}
\end{equation}

For an averaging interval $T_{\rm av}$, we define the time average of an observable $O(t)$ as
\begin{equation}
\overline{O}
=
\frac{1}{T_{\rm av}}
\int_0^{T_{\rm av}}
O(t)\,dt . \nonumber\\
\end{equation}

Taking the time average of Eq.~(\ref{eq:instant_sqvac}) gives
\begin{equation}
\overline{\Delta n^2}
=
2\overline{\langle n(t)\rangle^2}
+
2\overline{\langle n(t)\rangle}.
\label{eq:avg_sqvac_1}
\end{equation}

The temporal variance of the instantaneous mean occupation is defined as
\begin{equation}
{\rm Var}_t[\langle n(t)\rangle]
=
\overline{\langle n(t)\rangle^2}
-
\overline{\langle n(t)\rangle}^{\,2}. \nonumber\\
\label{eq:temporal_variance}
\end{equation}

Hence,
\begin{equation}
\overline{\langle n(t)\rangle^2}
=
\overline{\langle n(t)\rangle}^{\,2}
+
{\rm Var}_t[\langle n(t)\rangle].
\label{eq:mean_square_identity}
\end{equation}

Substituting Eq.~(\ref{eq:mean_square_identity}) into Eq.~(\ref{eq:avg_sqvac_1}) yields
\begin{equation}
{
\overline{\Delta n^2}
=
2\overline{\langle n\rangle}^{\,2}
+
2\overline{\langle n\rangle}
+
2\,{\rm Var}_t[\langle n(t)\rangle]
}.
\label{eq:avg_sqvac_final}
\end{equation}

Therefore, the relation between the time-averaged phonon-number variance and the time-averaged mean occupation contains an additional positive contribution arising from the temporal variation of $\langle n(t)\rangle$. In the RF-driven trap, this term includes the contribution of the micromotion and, for averages extending over many RF periods, any additional slower variation of the mean occupation. Consequently, the time-averaged quantities need not satisfy the instantaneous squeezed-vacuum relation evaluated simply at $\overline{\langle n\rangle}$.

\section{Extraction of the rotated-quadrature marginal from the Wigner function}
\label{app:marginal}

Panel (b) and (e) of Fig.~\ref{fig:wigner_combined} shows the probability distribution $P(X_\theta)$ of the rotated quadrature
\begin{equation}
  \hat X_\theta = \frac{\hat a\,e^{-i\theta}+\hat a^\dagger e^{i\theta}}{\sqrt2}
  = \hat X\cos\theta + \hat P\sin\theta , \nonumber\\
\end{equation}
which is a measurement axis in phase space oriented at an angle $\theta$ from
the $X$ axis. For a given state we obtain $P(X_\theta)$ directly from the Wigner
distribution $W(X,P)$ as its marginal along the direction orthogonal to
$\hat X_\theta$,
\begin{equation}
  P(X_\theta) = \int_{-\infty}^{\infty} W\!\left(X_\theta,\,X_{\theta+\pi/2}\right)
  \, dX_{\theta+\pi/2},
  \label{eq:radon}
\end{equation}
i.e.\ the projection (Radon transform) of $W$ onto the $X_\theta$ axis. This
marginal is the true probability distribution of the quadrature $\hat X_\theta$,
so its variance equals $\langle \hat X_\theta^2\rangle-\langle \hat X_\theta\rangle^2$
and thus determines $S_\theta$.

Numerically, $W(X,P)$ is evaluated on a uniform Cartesian grid
$\{(X_i,P_j)\}$ of spacing $\Delta$ covering the relevant region of phase
space. For a chosen angle $\theta$, each grid point is assigned its coordinate
along the rotated axis,
\begin{equation}
  u_{ij} = X_i\cos\theta + P_j\sin\theta , \nonumber\\
\end{equation}
which is the projection of $(X_i,P_j)$ onto the $X_\theta$ direction; all points
lying on a common line perpendicular to this axis share the same value of $u$.
The marginal is then built by binning the $u_{ij}$ into a one-dimensional
histogram of bin width $\delta u$ and accumulating the corresponding Wigner
weights,
\begin{equation}
  \tilde P(u_k) = \sum_{i,j}\, W(X_i,P_j)\,
  \Pi\!\left(\frac{u_{ij}-u_k}{\delta u}\right), \nonumber\\
\end{equation}
where $\Pi$ is the unit rectangle (bin) function and $u_k$ denotes the bin
centres. This sum collapses the two-dimensional distribution onto the
$X_\theta$ axis by adding the Wigner density stacked along the orthogonal
direction, thereby approximating the integral in Eq.~\eqref{eq:radon}. The
result is normalised to unit area,
\begin{equation}
  P(X_\theta=u_k) = \frac{\tilde P(u_k)}
  {\sum_{k'}\tilde P(u_{k'})\,\delta u}, 
\end{equation}
so that $P(X_\theta)$ is a genuine probability density. The squeezed and
anti-squeezed marginals in rows (b) and (e) correspond to $\theta=\theta_s$ and
$\theta=\theta_s+\pi/2$, respectively, and are compared with the vacuum
distribution $P_{\rm vac}(X_\theta)=\pi^{-1/2}e^{-X_\theta^2}$ of variance
$1/2$. A marginal narrower than this vacuum reference along $\theta_s$ signals
$S_\theta<0$, i.e.\ quadrature squeezing.

We note that the same $S_\theta$ follows equivalently, and without grid
discretisation, from the second moments evaluated directly from the propagated
Fock amplitudes; the Wigner projection described here is used for the
phase-space visualisation, and the two routes were verified to agree.

\bibliographystyle{apsrev4-2}
\bibliography{reference} 

\end{document}